# Multicomponent Gas Diffusion in Porous Electrodes

Yeqing Fu[1], Yi Jiang[2], Abhijit Dutta[2], Aravind Mohanram[2], John D. Pietras[2], Martin Z. Bazant[1,3]

[1] Department of Chemical Engineering, Massachusetts Institute of Technology, Cambridge, MA
[2] Saint-Gobain R&D Center, Northboro, MA
[3] Department of Mathematics, Massachusetts Institute of Technology, Cambridge, MA

## Abstract

Multicomponent gas transport is investigated with unprecedented precision by AC impedance analysis of porous YSZ anode-supported solid oxide fuel cells. A fuel gas mixture of $H_2$-$H_2O$-$N_2$ is fed to the anode, and impedance data are measured across the range of hydrogen partial pressure (10-100%) for open circuit conditions at three temperatures (800ºC, 850ºC and 900ºC) and for 300mA applied current at 800ºC. For the first time, analytical formulae for the diffusion resistance ($R_b$) of three standard models of multicomponent gas transport (Fick, Stefan-Maxwell, and Dusty Gas) are derived and tested against the impedance data. The tortuosity is the only fitting parameter since all the diffusion coefficients are known. Only the Dusty Gas model leads to a remarkable data collapse for over twenty experimental conditions, using a constant tortuosity consistent with permeability measurements and the Bruggeman relation. These results establish the accuracy of the Dusty Gas model for multicomponent gas diffusion in porous media and confirm the efficacy of electrochemical impedance analysis to precisely determine transport mechanisms.

# I. Introduction

The Solid Oxide Fuel Cell (SOFC) is currently the highest-temperature fuel cell in development and can be operated over a wide temperature range from 600ºC-1000ºC allowing a number of fuels to be used. To operate at such high temperatures, the electrolyte is a thin, nonporous solid ceramic membrane that is conductive to charge carrier, $O^{2-}$ ions. The operating efficiency in generating electricity is among the highest of the fuel cells at about 60%[1]. Furthermore, the high operating temperature allows cogeneration of high-pressure steam that can be used in many applications. Combining a high-temperature SOFC with a turbine into a hybrid fuel cell further increases the overall efficiency of generating electricity with a potential of an efficiency of more than 70%[1]. Therefore, it is a very promising alternative energy source that could potentially be used for home heating or large scale electricity production in the future.

Solid oxide fuel cell consists of a porous cathode, an electrolyte, a porous anode and interconnects. Two different types have been explored in the development of SOFC, the electrolyte supported cell and the electrode supported cell. In the former, electrolyte is the thickest and serves as the mechanical support for the whole cell. However, due to the high Ohmic resistance of the relatively thick electrolyte layer, the electrolyte supported design has been gradually replaced by the new electrode supported cells, in which one of the porous electrodes is the supporting structure. Moreover, since cathode supported cell usually gives higher resistance, and is much harder to fabricate due to the mismatched thermal expansion coefficient of cathode support and functional layer, the anode supported cell (ASC) is the most widely accepted design in current SOFC research.

The solid oxide fuel cell is operated with fuel and oxidant being continuously fed from two sides of the cell. Fuel (typically, hydrogen and/or hydrocarbon mixture) is provided to the anode side while oxygen carried by air is provided to the cathode. As the fuel and air react, water vapor is produced and removed from anode. Fuels and oxidants have to be transported through porous electrodes before they arrive at the functional layer, the reaction site. At the same time, product or water vapor has to travel through the porous anode in the opposite direction to be taken away by the flowing stream. Therefore, gas transport through the porous electrode is an essential factor determining the overall cell performance[2,3]. The efficacy of the gas transport through the porous electrodes often determines the rate of electrochemical reaction or current generation.

Furthermore, many researches have shown that the gas transport through porous electrodes is mainly governed by gas diffusion with very small convection contribution[4–7]. Thus, gas diffusion in porous electrodes is the main source of concentration polarization (concentration difference between bulk gas and functional layers) in solid oxide fuel cells. However, the diffusion process has not been well understood yet due to 1) gas phase is a multicomponent gas mixture, including reactants, carrier gas and possibly products; 2) the porous electrode, through which gas phase has to travel, could have very complicated microstructures.

There is abundant literature on modeling gas diffusion in porous media using Fick's law, Stefan-Maxwell or Dusty Gas model. It is thought that Dusty Gas model should be the most accurate, although it is also the most complicated and difficult to validate. Almost no analytical results are available, but the Dusty Gas model has been used in a number of numerical simulations[2–4,8,9], albeit with constant pressure approximation which is inconsistent[9] (see below). Moreover, no theoretical framework exists to analytically derive the diffusion resistance values from impedance data using these more complex diffusion models for porous media.

Instead, current researchers usually use limiting current values from the current-voltage or I-V curves to study gas diffusion in SOFC[10]. Limiting current is usually obtained when the reactant is nearly or completely depleted at the reaction site. Therefore, it has often been used to derive properties of the porous electrode that would account for slow diffusion or sluggish mass transport[10–12]. However, high tortuosity (ratio of actual distance travelled by gas to straight line distance between two points) is commonly invoked to explain the limiting current values. Many previous attempts to fit models to I-V data for SOFC have been inconclusive with widely varying tortuosity values from 2 to 19 for the same system[13–16]. Yet, most direct measurements conducted on anode materials and reconstruction of 3D microstructure[17–20] indicate tortuosity values should be in the range of 1.5 to 3. At the same time, according to the theories about tortuosity[21], we should expect tortuosity of porous electrode with nice and open microstructures to be not too high. Actually, limiting current can have the signature of not only gas diffusion[22], but also dissociative adsorption, surface diffusion, catalytic redox reaction, or even gas transport in free channels outside the electrode. Therefore, we studied gas diffusion in porous electrodes using AC impedance, which better separates processes of different time scales and therefore provides better assignment of arcs in data to different processes.

The SOFC button cell we studied uses hydrogen as fuel, carried by nitrogen together with 1.7% of water vapor, for anode. Oxygen in air is used as oxidant for cathode. Therefore, the electrochemical reaction goes as follows. Oxygen molecule diffuses through cathode bulk layer and reaches the functional layer, where it accepts electrons and is oxidized to oxygen ion, which is further conducted through the electrolyte layer. When it arrives at the anode functional layer, it reacts with hydrogen fuel, forming water and releasing electrons to the external circuit.

In this paper, we present a new theoretical approach to predict concentration profiles and diffusion resistance using Fick's law (Fick), Stefan-Maxwell formulation (SM), and Dusty Gas model (DGM) and compare with experimental data for SOFC. By using this approach in conjunction with AC impedance, we are able to show that DGM provides a very accurate description of multicomponent gas diffusion and can be used to subtract gas diffusion response from overall data for analyzing contributions from other physical processes.

**II. Theory**

**1. Models**

Transport of gaseous components through porous media has been extensively studied over the years. In general, models including Fick's model (FM), the Stefan–Maxwell model (SMM) and the Dusty-gas model (DGM) are widely used to predict the concentration overpotential. Many researches have concluded that among the above three, the dusty-gas model is the most accurate and appropriate model to simulate gas transport phenomena inside a porous electrode[4,6,8], such as SOFC electrodes. However, due to its complexity, this model has no analytical solutions, and the corresponding analysis requires complicated numerical simulation[2,3,6,23–25]. In this work, we developed a new theoretical approach which is based on impedance analysis to show how DGM can also been used to analytically analyze the gas diffusion inside the porous media.

Fick's law is the simplest diffusion model and is used in dilute or binary systems. It assumes the net flux is proportional to the gradient of the concentration of the corresponding species[26].

$$N_i = -\frac{P}{RT} \cdot D_i^{eff} \frac{dX_i}{dx} \quad (1)$$

$D_i^{eff}$ in Fick's law is the effective diffusion coefficient of species $i$, which takes into account of the composition of the gas mixture. The calculation of $D_i^{eff}$ can be carried out following equation 2. Where $D_i$ is the theoretical diffusion coefficient of species $i$, $\varepsilon_P$ and $\tau_P$ are the porosity and tortuosity of the porous electrode, respectively.

$$D_i^{eff} = D_i \frac{\varepsilon_P}{\tau_P} \quad (2)$$

Stefan-Maxwell model is more commonly used in multi-component systems because it considers the molecular collisions among different types of the gas species by using a more complicated left hand side term (equation 3)[27–29]. However, it is more typically used for nonporous media. In equation 3, $X_i$ is the mole fraction of species $i$ in the gas mixture, $N_i$ is the mole flux of species $i$, $P$ is total gas pressure in Pa, $R$ is the universal gas constant, $T$ is absolute temperature in K, and $x$ is the 1 D spatial variable.

$$\sum_{j \neq i} \frac{X_j N_i - X_i N_j}{D_{i,j}^{eff}} = -\frac{P}{RT} \frac{dX_i}{dx} \quad (3)$$

The Dusty Gas model is an extension of the Stefan-Maxwell equation. It further takes into account the molecules-pore wall interactions by introducing the Knudsen diffusion term (first term in eqn. 4)[11,25]. This model assumes the pore walls consist of large molecules that are uniformly distributed in space. These pseudo or dummy 'dust' molecules also collide with real gas molecules, bringing in the Knudsen diffusion effect.

Besides, the viscous fluxes due to pressure gradient are also taken into consideration. The general form of the DGM is the following (Equation 4)

$$\frac{N_i}{D_{K,i}^{eff}} + \sum_{j \neq i} \frac{X_j N_i - X_i N_j}{D_{i,j}^{eff}} = -\frac{P}{RT}\frac{dX_i}{dx} - \frac{X_i}{RT}\left(1 + \frac{1}{D_{K,i}^{eff}}\frac{B_0 P}{\mu}\right)\frac{dP}{dx} \quad (4)$$

where $B_0$ is the permeability of the porous medium and $\mu$ is the viscosity of the gas mixture.

In both SM and DGM, the binary diffusion coefficients $D_{i,j}$ can be calculated by the Chapman-Enskog equation (Eqn.5), where T is temperature in K, $p$ is pressure in Pa, $\Omega$ is the collision integral, $\sigma_{ij}$ is the collision diameter, and $M_i$ is the molecular weight of species $i$ [31].

$$D_{i,j} = \frac{1.86 \times 10^{-3} T^{3/2} \left(\frac{1}{M_i} + \frac{1}{M_j}\right)^{1/2}}{p \Omega \sigma_{ij}^2} \quad (5)$$

Knudsen diffusion coefficients can be derived from the kinetic theory of gases (equation. 6) where $\bar{r}$ is the radius of the gas molecule, $M_i$ is the molecular weight of species $i$ [31].

$$D_{K,i} = \frac{2}{3}\left(\frac{8RT}{\pi M_i}\right)^{1/2} \bar{r} \quad (6)$$

Note that this expression was derived from cylindrical pore geometry that having the mean radius, but in reality, pore geometry can deviate from cylinders, therefore, this expression has some uncertainties in predicting Knudsen diffusivity.

The effective binary diffusivity and Knudsen diffusivity ($D_{i,j}^{eff}$ and $D_{K,i}^{eff}$) were defined as their theoretical counterparts ($D_{i,j}$ and $D_{K,i}$) times a geometric factor, which is porosity ($\varepsilon_P$) divided by tortuosity ($\tau_P$).

$$D_{i,j}^{eff} = D_{i,j}\frac{\varepsilon_P}{\tau_P} \quad and \quad D_{K,i}^{eff} = D_{K,i}\frac{\varepsilon_P}{\tau_P} \quad (7)$$

Numerous studies on transport through porous media in the absence of a chemical reaction reveal that the DGM is superior to the Fick's law in its ability to predict the fluxes[32,33]. In porous catalyst, the Fick's law is still frequently used because its simplicity allows explicit and analytical expressions to be derived. If nonuniform pressure is present in a porous media due to reactions involving a change in the number of molecules, additional permeation term has to been taken into account, and therefore DGM should be adopted. Many works[34,35] showed that the DGM can successfully predict the fluxes for these reactions in various reacting systems. For example, Davies[36] used it for the $SO_2$ oxidation reaction, Bliek[37] applied it to the coal gasification where large pressure gradient is present, However, the pressure gradient term requires additional computational time and cost. Therefore researchers started to use DGM without the permeation term if pressure gradient can be approximately neglected. And a comparison among different diffusion models, including Fick's Law, Stefan-Maxwell model and Dusty Gas Model, to predict concentration polarization is presented in in Suwanwarangkul' work[4].

**Debates on Graham's law:**

Interestingly, we found there is a paradox of Dusty Gas model with constant pressure assumption, which has not been widely realized in the community of SOFC. Eqn.4 shows the general Dusty Gas model with an extra permeation flux term due to the pressure variation, if we sum over all the gas species, the pressure gradient can be calculated as shown in Eqn.9. By taking a look at the numerator, we can find that the pressure gradient comes from the different effective Knudsen diffusivity $D_{K,i}^{eff}$ of two active species in equi-molar counter-diffusion mode.

$$\sum_i \frac{N_i}{D_{K,i}^{eff}} = -\frac{1}{RT}\left(1 + \frac{B_0 P}{\mu}\sum_i \frac{X_i}{D_{K,i}^{eff}}\right)\frac{dP}{dx} \quad (8)$$

$$\frac{dP}{dx} = \frac{-RT\sum_i \frac{N_i}{D_{K,i}^{eff}}}{\left(1 + \frac{B_0}{\mu}\sum_i \frac{P \cdot X_i}{D_{K,i}^{eff}}\right)} \quad (9)$$

In the case of hydrogen molecules reacting to produce water vapor, the molar flux of all species should add up to zero. In this equi-molar counter diffusion mode, if the effective Knudsen diffusivity $D_{K,i}^{eff}$ of hydrogen and water are the same, which means if the force exerted on the pore walls by $H_2$ and $H_2O$ are exactly the same but in the opposite direction, they will cancel each other and no pressure will build up. However, the molecular weight and size of the molecules vary among different species, therefore, Knudsen diffusivity must be different, which means total pressure has to change throughout the electrode.

From another point of view, in the constant pressure assumption, the summation over all gas components will lead to Graham's law[38], which says the sum of molar flux ($N_i$) times the square root of the molecular weight ($M_i$) should add up to zero (Eqn.10). Actually, the Graham's law is valid in the absence of chemical reactions. But when chemical reactions occur, the component fluxes are related through the reaction stoichiometry, and only isomerization reactions will be consistent with Graham's law.

$$\sum_i N_i \sqrt{M_i} = 0 \quad (10)$$

In our case, moles of $H_2$ react to form equivalent number of moles of $H_2O$ and this is obviously contradictory to the flux relations imposed by the reaction. Since the algebraic derivation from Dusty Gas model to Graham's law is strict, this conflict indicates the Dusty Gas Model is intrinsically inconsistent with the constant pressure assumption. Actually, Graham's law is only valid in the case of gas diffusion without reaction in general. In the case of SOFC, the gas diffusion in porous electrode has a boundary condition of surface reaction at the functional layer/electrolyte interface; therefore, the flux of active species ($H_2$ and $H_2O$) cannot be captured by Graham's law. However, some current researches still use it to study gas transport in porous SOFC electrodes[4,39]. In fact, full DGM with permeation flux term due to pressure variation has no problem, and is accurate enough to satisfy chemical reaction boundary conditions. Yet with the permeation term, DGM is too complicated for deriving analytical results, therefore restrict its acceptability in some theoretical studies. But we will provide a proof, in the section II.4 that in porous electrodes of SOFC, the pressure gradient effects on the gas transport is not significant and can be safely neglected.

## 2. Steady State Concentration Profiles

From the governing equations, we can derive the concentration profiles throughout the porous electrode when the bulk concentrations of different gas species are taken to be known. In Eqns. 11-13, $X_i^0$ is the molar fraction of species $i$ in the bulk gas mixture outside the porous electrode, $X_i(x)$ is the molar fraction of species $i$ at position $x$. $I$ is the total current, $F$ is the Faraday constant. $R$ is universal gas constant and $P$ is the total gas pressure.

$$X_{N2}(x) = X_{N2}^0 \exp\left[\frac{RTI}{2FP}\left(\frac{1}{D_{H2,N2}^{eff}} - \frac{1}{D_{H2,H2O}^{eff}}\right)x\right] \quad (11)$$

$$X_{H2} = X_{H2}^0 - \frac{RTI}{2FP}\left(\frac{1}{D_{K,H2}^{eff}} + \frac{1}{D_{H2,H2O}^{eff}}\right)x - \left(\frac{(D_{H2,H2O}^{eff} - D_{H2,N2}^{eff})D_{N2,H2O}^{eff}}{(D_{N2,H2O}^{eff} - D_{H2,N2}^{eff})D_{H2,H2O}^{eff}}\right)X_{N2}^0\left\{\exp\left[\frac{RTI}{2FP}\left(\frac{1}{D_{H2,N2}^{eff}} - \frac{1}{D_{N2,H2O}^{eff}}\right)x\right] - 1\right\} \quad (12)$$

$$X_{H2O} = X_{H2O}^0 + \frac{RTI}{2FP}\left(\frac{1}{D_{K,H2O}^{eff}} + \frac{1}{D_{H2,H2O}^{eff}}\right)x + \left(\frac{(D_{H2,H2O}^{eff} - D_{H2O,N2}^{eff})D_{N2,H2}^{eff}}{(D_{N2,H2O}^{eff} - D_{H2,N2}^{eff})D_{H2,H2O}^{eff}}\right)X_{N2}^0\left\{\exp\left[\frac{RTI}{2FP}\left(\frac{1}{D_{H2,N2}^{eff}} - \frac{1}{D_{N2,H2O}^{eff}}\right)x\right] - 1\right\} \quad (13)$$

### Concentration Polarization

From the concentration profile calculation, we know the gas concentration at the reaction surface and then concentration overpotential can be calculated using Nernst equation (Eqn.14)

$$\eta_{anode\_conc} = -\frac{RT}{2F}\ln\left(\frac{X_{H_2} X_{H_2O}^0}{X_{H_2}^0 X_{H_2O}}\right) \quad (14)$$

In the case of impedance under current, the concentration outside the porous electrode is very close to the bulk concentration (the concentration in the feed gas). However, under a non-zero current, some of the reactants need to react electrochemically to support the current, therefore, there must be some concentration gradient resulting from the consumption of the reactants. We use a continuously stirred tank reactor (CSTR) assumption to approximately calculate the gas concentration outside the cell in the feed tube[16], as described in Eqn.20 and Eqn.21.

### 3. Diffusion impedance ($R_b$) (with and without dP)

The above mentioned gas diffusion models, including Fick's law, Stefan Maxwell and Dusty Gas model, are not new, and are widely used to predict I-V curves and fit the limiting current values as mentioned earlier[13,20]. But the SM and DGM models have rarely been used before to analytically analyze impedance spectra of SOFC, although they have been used to describe gas diffusion in porous electrodes. By taking the derivative of the concentration overpotential with respect to current and evaluating it at a specified current, diffusion resistance ($R_b$) is obtained for all three models. By taking a look at zero current $R_b$ in equations.16, 17, and 18, we notice that compared to $R_b$ value from Fick's law, the $R_b$ of SM has an extra complicated term resulting from the consideration of interactions among different gas species. Also, the $R_b$ value derived from DGM further incorporated the Knudsen effect, which accounts for the collision of gas molecules with the pore wall. The multicomponent gas diffusion inside the porous electrodes was then studied by comparing these three different diffusion models. In Eqns. 15 to 18, $R_{b\_anode(I)}$ is the gas diffusion resistance at current I, $\eta_{anode}$ is the anode concentration overpotential due to gas diffusion, $P_i^0$ and $X_i^0$ is the partial pressure and molar fraction of species i in the bulk gas mixture outside the porous electrode. All other parameters are defined the same way as in general Dusty Gas model.

$$R_{b\_anode(I=0)} = \frac{d\eta_{anode}}{dI}\bigg|_{(I=0)} \text{ or } R_{b\_anode(I)} = \frac{d\eta_{anode}}{dI}\bigg|_{(I)} \quad (15)$$

$$Rb_{Fick(anode)} = \left(\frac{RT}{2F}\right)^2 L_a \left\{ \frac{1}{P_{H2O}^0}\left(\frac{1}{D_{K,H2O}^{eff}} + \frac{1}{D_{H2,H2O}^{eff}}\right) + \frac{1}{P_{H2}^0}\left(\frac{1}{D_{K,H2}^{eff}} + \frac{1}{D_{H2,H2O}^{eff}}\right) \right\} \quad (16)$$

$$Rb_{SM(anode)} = \left(\frac{RT}{2F}\right)^2 L_a \left\{ \begin{array}{l} \frac{1}{P_{H2O}^0}\frac{1}{D_{H2,H2O}^{eff}} + \frac{1}{P_{H2}^0}\frac{1}{D_{H2,H2O}^{eff}} \\ + X_{N2}^0 \left(\frac{1}{D_{H2,N2}^{eff}} - \frac{1}{D_{N2,H2O}^{eff}}\right) \cdot \left(\frac{\left(D_{H2,H2O}^{eff} - D_{H2O,N2}^{eff}\right)D_{N2,H2}^{eff}}{\left(D_{N2,H2O}^{eff} - D_{H2,N2}^{eff}\right)D_{H2,H2O}^{eff}P_{H2O}^0} + \frac{\left(D_{H2,H2O}^{eff} - D_{H2,N2}^{eff}\right)D_{N2,H2O}^{eff}}{\left(D_{N2,H2O}^{eff} - D_{H2,N2}^{eff}\right)D_{H2,H2O}^{eff}P_{H2}^0}\right) \end{array} \right\} \quad (17)$$

$$Rb_{DGM(anode)} = \left(\frac{RT}{2F}\right)^2 L_a \left\{ \begin{array}{l} \dfrac{1}{P^0_{H2O}}\left(\dfrac{1}{D^{eff}_{K,H2O}} + \dfrac{1}{D^{eff}_{H2,H2O}}\right) + \dfrac{1}{P^0_{H2}}\left(\dfrac{1}{D^{eff}_{K,H2}} + \dfrac{1}{D^{eff}_{H2,H2O}}\right) \\ + X^0_{N2}\left(\dfrac{1}{D^{eff}_{H2,N2}} - \dfrac{1}{D^{eff}_{N2,H2O}}\right) \cdot \left(\dfrac{\left(D^{eff}_{H2,H2O} - D^{eff}_{H2O,N2}\right)D^{eff}_{N2,H2}}{\left(D^{eff}_{N2,H2O} - D^{eff}_{H2,N2}\right)D^{eff}_{H2,H2O}P^0_{H2O}} + \dfrac{\left(D^{eff}_{H2,H2O} - D^{eff}_{H2,N2}\right)D^{eff}_{N2,H2O}}{\left(D^{eff}_{N2,H2O} - D^{eff}_{H2,N2}\right)D^{eff}_{H2,H2O}P^0_{H2}}\right) \end{array} \right\} \quad (18)$$

Note that in the $R_b$ expressions, there are not too many quantities that need to be fitted to data. Almost all the variables and parameters are determined from experimental inputs or estimations from kinetic gas theory, except for a microstructure factor (porosity divided by tortuosity), which links effective diffusivity inside porous electrode with its theoretical value. When the porosity is known, the only quantity need to be determined from fitting is the tortuosity value of the electrodes.

Similarly, cathode diffusion resistance can also be estimated by deriving from a specified diffusion model, e.g., Dusty Gas model derivation was shown in Eqn.19

$$Rb_{DGM(cathode)} = -\left(\frac{RT}{4F}\right)^2 \frac{L_c}{P^0_{O2}}\left(\frac{1}{D^{eff}_{K,O2}} + \frac{X^0_{N2}}{D^{eff}_{N2,O2}}\right) \quad (19)$$

Comparing the theoretical $R_b$ at anode and cathode, Figure.1 shows the ratio between cathode $R_b$ and anode $R_b$ multiplied by 100%. It clearly shows that theoretical $R_b$ of cathode is less than 0.5% of that of the anode in anode supported cells. Though the porosity and tortuosity can be slightly different in two porous electrodes, we can still safely conclude that the diffusion resistance from anode side dominates. Therefore, in all the following discussion, we treat total gas diffusion resistance to be anode gas diffusion resistance, and the low frequency arc in the impedance data was fit with a finite-length Warburg element in a Randles circuit (Fig.2) to extract the anode diffusion resistance ($R_b$), which was then compared to analytical predictions from the three diffusion models (Eqns.16,17,18).

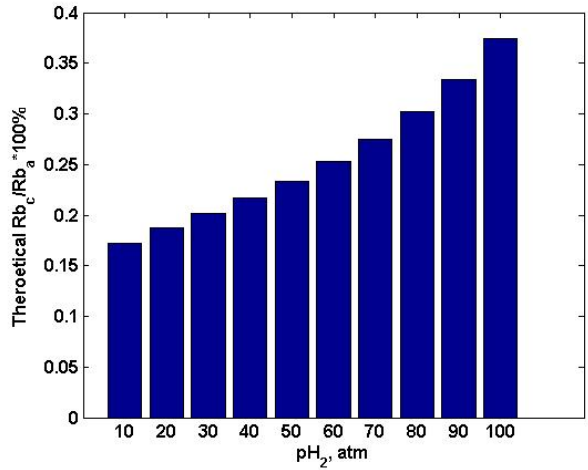

Fig. 1 Theoretical comparison of gas diffusion resistance ($R_b$) from cathode and anode in anode supported cell at different $pH_2$ levels.

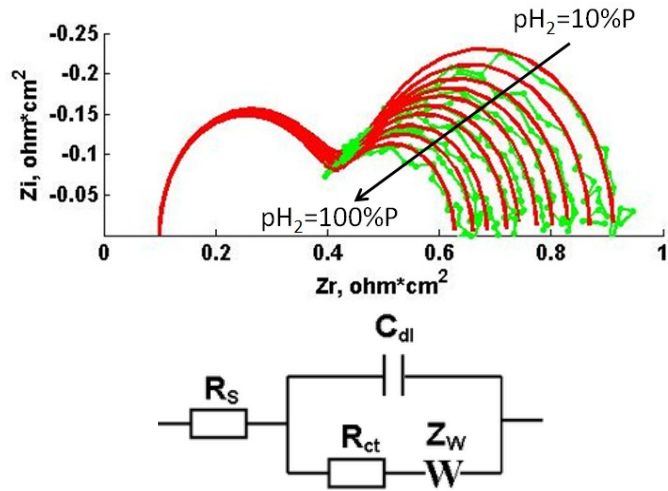

Fig.2 Fitting (top) of the low frequency arcs with the Warburg element in a Randles circuit (bottom).

**Nonlinear least-squares fitting (CNLS)**

A fitting procedure called complex nonlinear least-squares fitting (CNLS), was implemented, where data sets of ($Z_{real}$, $Z_{imaginary}$) versus frequency, or ($|Z|$, phase angle) versus frequency $\omega$ are used. The aim of the least squares fitting procedure is to find a set of parameters which will minimize the sum of weighted deviations.

$$\sum_{k=1}^{n} w_k \cdot \left[ (Z_{k,r} - Z'_{k,r}(\omega))^2 + (Z_{k,i} - Z'_{k,i}(\omega))^2 \right]$$

Where subscript $k$ denotes the $k$ th data point in impedance plot, $Z_{k,r}$ is the real part of the experimental impedance data, while its counterpart $Z'_{k,r}$ is the theoretical prediction of the real part of the impedance response. Similarly, $Z_{k,i}$ and $Z'_{k,i}$ are the imaginary parts of the impedance, experimentally and theoretically, respectively. Note that the theoretical prediction of the impedance is a function of frequency $\omega$, which makes fitting of the Nyquist plot to be a three dimensional curve fitting. Curves should not only match the correspondence of real and imaginary parts, but also need to match their frequency dependence as well. $w_k$ is the weighting factor, for which we use the magnitude of the $k$ th data point in this study. By minimizing the sum using the least square logarithm, a set of optimized parameters will be obtained.

We chose a Levenberg–Marquardt nonlinear least-squares fitting algorithm because of its straightforward implementation. Any parameter entering the model can in principle be used as a free fitting parameter; however, care must be taken to stay within limits of physical sense. It should also be noted that the Levenberg–Marquardt algorithm does not necessarily iterate to a global optimum of the fitting parameters, nor does it give any indication for the uniqueness of the optimized parameters. Therefore it is important to start from realistic initial guesses for the free parameters, and to exclude fitted results by analyzing its orders of magnitude and looking at the fitted graphs. Or upper and lower bounds can be set in the process of nonlinear least square fitting.

### 4. Proof of using isobaric assumption in Dusty Gas model

As mentioned before, Dusty Gas model is intrinsically inconsistent with isobaric or constant pressure assumption. However, by comparing the gas composition profiles of $H_2$, $N_2$ and $H_2O$, we can see pressure variation only leads to very small deviations of the gas composition profile (Fig.3). At the same time, theoretical prediction of anode gas diffusion resistance $R_b$ values derived from the full Dusty Gas model is practically the same as that derived from the isobaric Dusty Gas model (Fig.4). Therefore, our analysis proved that it is still safe to neglect total pressure variation inside the porous electrode when using the Dusty Gas model, although theoretically there is some inconsistency between the model itself and the isobaric assumption. Therefore, all the analysis and results we show in this paper are based on isobaric assumption, assuming total pressure inside porous electrode does not vary in depth.

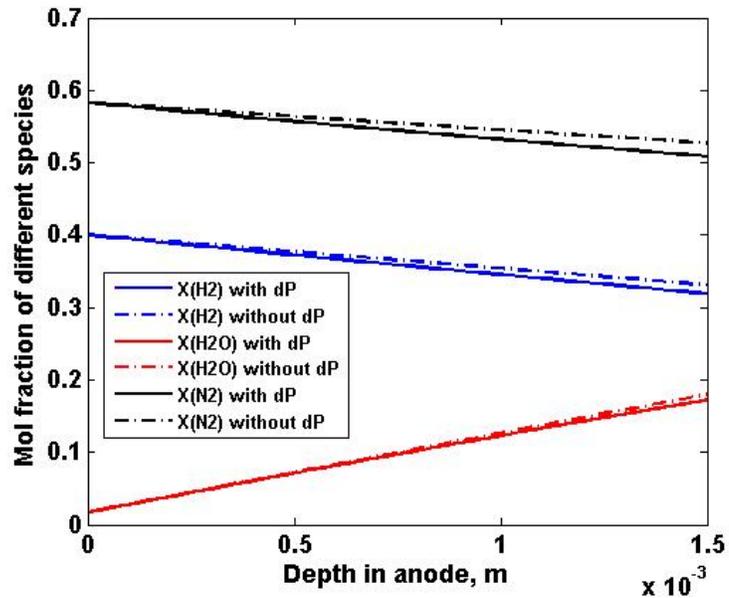

Fig. 3 Comparison of anode gas composition profiles under the current of I=100mA (full Dusty Gas model versus isobaric Dusty Gas model)

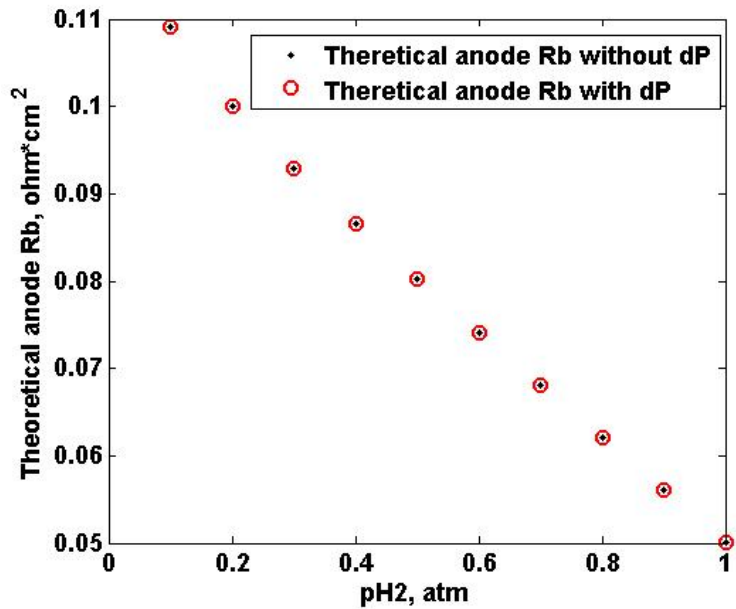

Fig. 4 Comparison of theoretical prediction of anode gas diffusion resistance ($R_b$) derived using full Dusty Gas model versus isobaric Dusty Gas model.

III. **Experiments**

Anode-supported single cells were fabricated based on technology developed by Forschungszentrum Jülich. The anode was approximately 1.5 mm thick and 1 inch in diameter, while the cathode was 0.1 mm thick, and 0.5 inch in diameter. The anode side consists of an anode support layer and an anode functional layer with a thickness of 15~30 um, both of which are composed of Ni/YSZ composites with different loadings and microstructures. The cathode side consists of a cathode current collection layer made from pure LSM and a cathode functional layer with a thickness of 15~30 um, which is composed of YSZ/LSM composite.

A new cell was sealed at the circumference using LP-1071 glass from Applied technologies and dried in an oven at $120^{o}C$ for 20 min. Then, it was placed into a spring loaded single cell testing fixture. The fixture was put into a furnace with $N_2$ (150sccm) on the anode side and Air (150sccm) on the cathode side. The furnace was then heated to $800^{o}C$ at $5^{o}C/min$. The cell was reduced the next morning for 3 hours by gradually switching the anode gas from $N_2$ to $H_2$ flowing at 300 sccm. During testing, a tertiary gas mixture of hydrogen, nitrogen and 3% by volume water vapor was provided to the anode from a top feeding tube and air was fed from the bottom, carrying oxygen to the cathode. Pt lead wires were connected from the current collector layers (Pt mesh on cathode and nickel on the anode) to the data collecting equipment. A 1470E Solartron Analytical from MTechnologies and mSTAT program were used to control the operating conditions and collect the data.

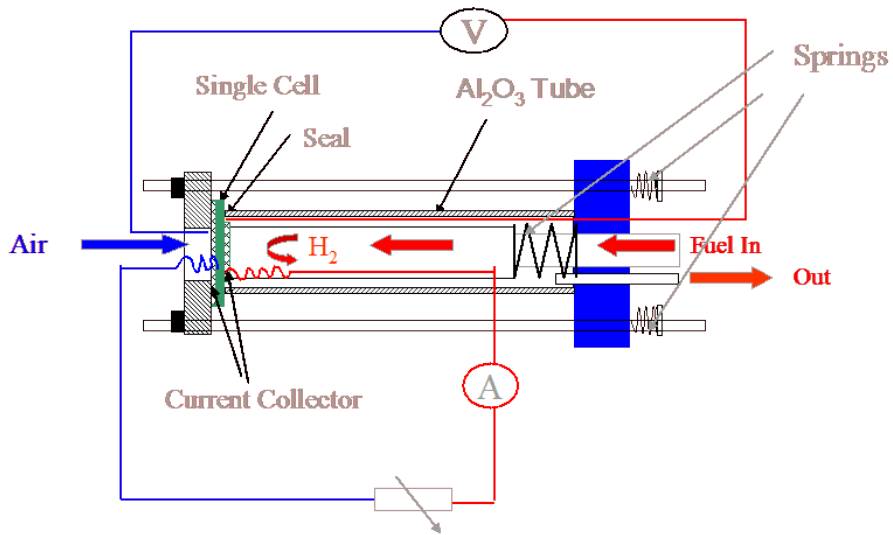

Fig. 5 The setup of the spring loaded testing fixture that used for anode supported cell testing.

AC impedance data of anode-supported cell was collected at open circuit voltage (OCV) condition (Fig. 6) and 300 mA current (Fig. 10) when varying the hydrogen partial pressure from 10% to 100% of the total pressure. OCV Impedance data were also collected at three different temperatures (800ºC, 850ºC, and 900ºC) (Fig. 12).

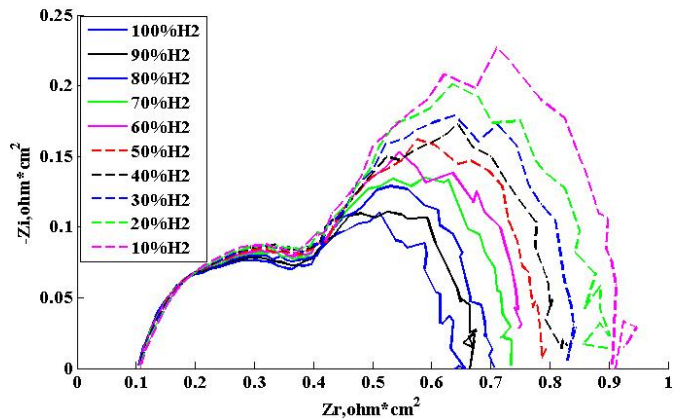

Fig. 6 OCV AC impedance data of anode supported cell collected at various hydrogen partial pressures. Total pressure of the anode feeding gas was fixed at 1 atm.

## IV. Analysis and Results

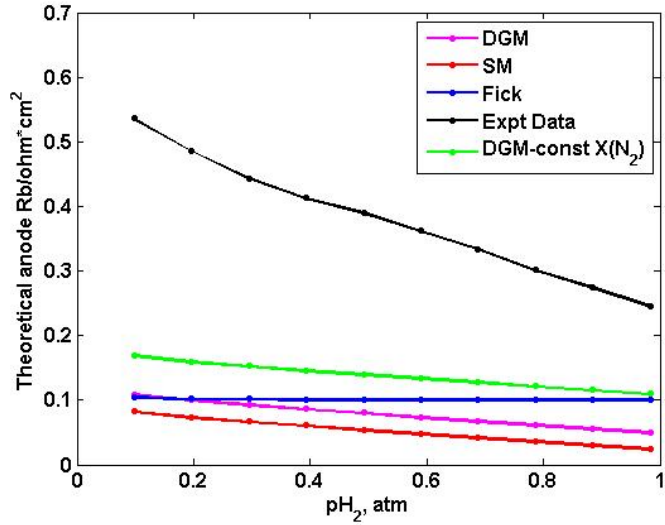

Fig. 7 Experimentally extracted anode $R_b$ and theoretical predictions of anode $R_b$ of anode supported cell under different hydrogen partial pressures at OCV.

Figure 7 shows the experimental $R_b$ values we extracted from experimental data (black curve), and the theoretical predictions of anode $R_b$ without taking into account the microstructure factor from three different models (colored curves). From the fitting, it is noteworthy that the Dusty Gas model gives a constant structural factor (porosity divided by tortuosity), independent of hydrogen partial pressure (Fig.8). This is consistent with real physics, where the microstructure of the porous media does not change with testing conditions. Moreover, with the anode porosity known to be 46%, the tortuosity fitted from the Dusty Gas model is 2.30, which matches both theoretical expectations and direct experimental measurements. After taking into account the fitted tortuosity, the Dusty Gas model best describes the gas diffusion, while the Stefan-Maxwell model shows some deviations, and Fick's law cannot capture the performance at all (Fig.9). It is also interesting to note that Dusty Gas model with constant $N_2$ composition does not give good enough results as well, which confirmed the necessity to calculate nitrogen concentration without any assumption, and use it to further calculate concentration profiles of other active species, such as $H_2$ and $H_2O$.

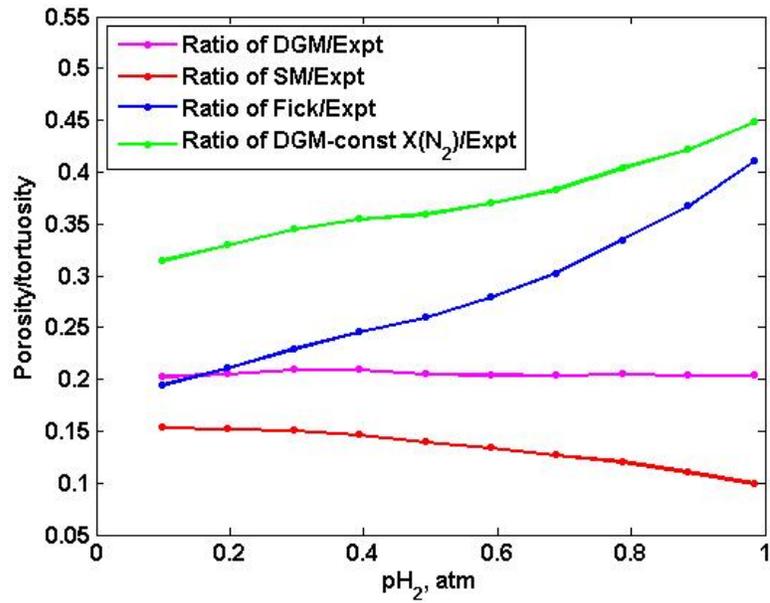

Fig.8 Structural factor (porosity/tortuosity) values fitted from three diffusion models under OCV.

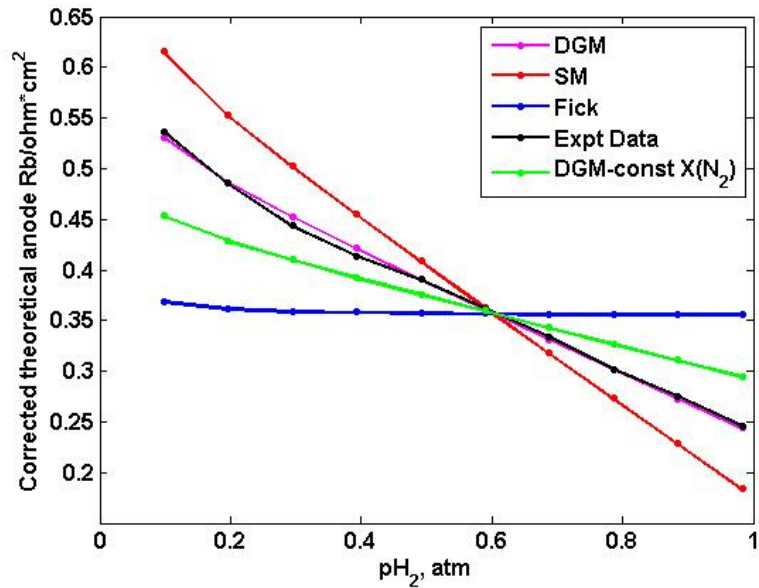

Fig.9 Comparison between diffusion resistances ($R_b$) derived from models and the values extracted from experimental data after taking into account the fitted structural factor (porosity/tortuosity).

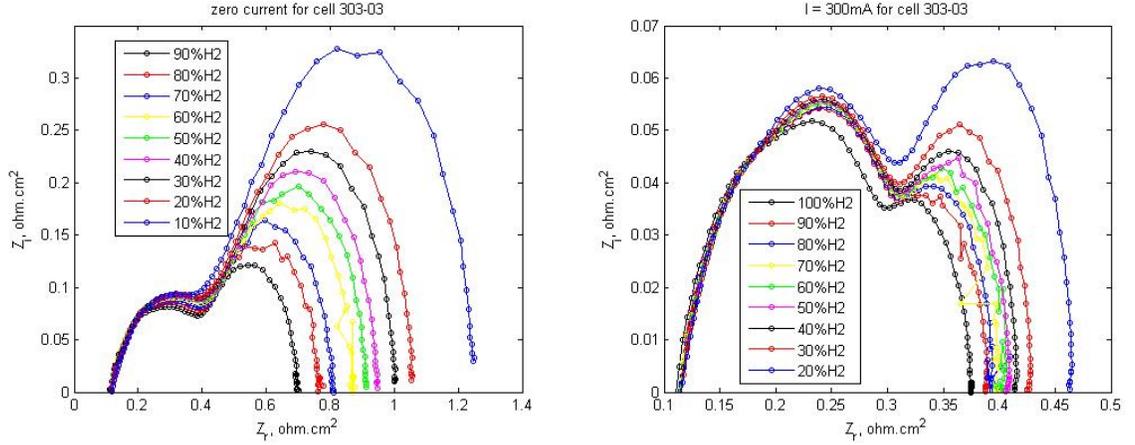

Fig. 10 AC impedance data of anode supported cells for anode supported cell (303-03) at OCV (left) and under a current of 300mA, when hydrogen partial pressure is varied.

We also derived the diffusion resistance $R_b$ for a non-zero current impedance. The impedance was measured at the current of 300 mA/cm$^2$ (Fig.10), and the corresponding $R_b$ values for all three models were numerically evaluated using Maple software. It is worth noting that at a non-zero current, the bulk gas concentration (concentration at the interface between porous electrode and gas feeding tube) can deviate from the feeding concentration due to the concentration polarization resulting from consumption of reactants by electrochemical reactions. And, the concentration gradients in the gas feeding tube can be approximated using a continuous stirred tank reactor (CSTR) model. Therefore a CSTR correction (Eqn.20, 21) is introduced for calculating the concentration boundary condition of the bulk gas concentration. $P_i^0$ is the ideal bulk concentration of species $i$, and $P_i^{0*}$ is the corrected bulk concentration of species $i$ after the CSTR formulation. In these equations, $N_i$ is the molecular flux of species $i$ in $mol/(m^2 \cdot s)$, $A$ is the electrode area in m$^2$, $m_T$ is the total flux of feeding gas in $mol/s$

$$P_{H_2}^{0*} = P_{H_2}^0 - \frac{N_{H_2} A}{m_T} P \quad (20)$$

$$P_{H_2O}^{0*} = P_{H_2O}^0 + \frac{N_{H_2O} A}{m_T} P \quad (21)$$

With the CSTR correction, the fitted porosity/tortuosity is almost independent of feeding gas composition and is practically the same as the zero current impedance analysis, which is another validation of the method of analyzing $R_b$ to extract tortuosity.

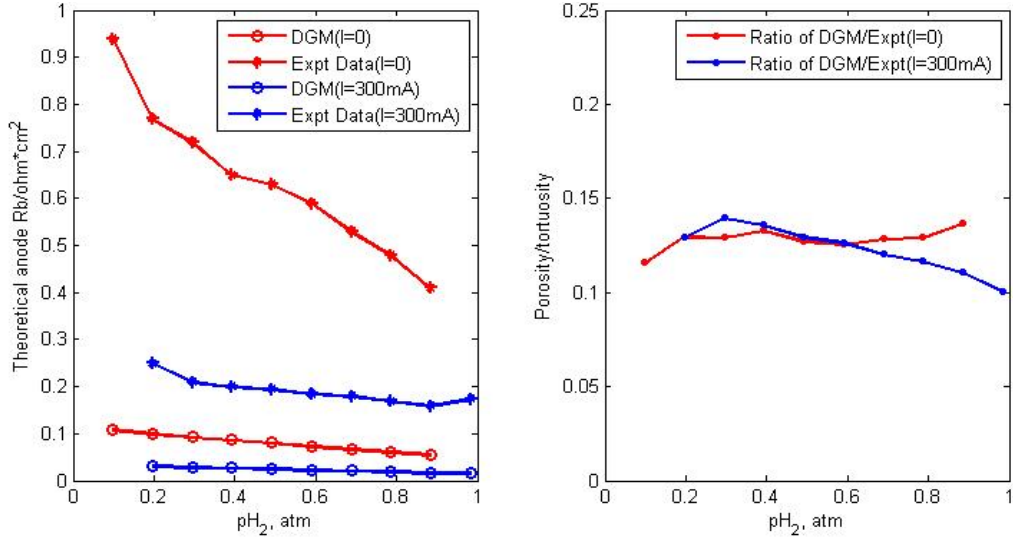

Fig. 11 Experimentally extracted anode $R_b$ and theoretical predictions for anode $R_b$ of anode supported cell under different hydrogen partial pressures at both zero and 300 mA/cm$^2$ current, using Dusty Gas model (left). Comparison of the fitted structure factors at two different currents (right).

We further applied this gas diffusion analysis to zero-current impedance measured at different temperatures (800ºC, 850ºC and 900ºC) (Fig. 12), and the tortuosity fitted at these temperatures only varies a little, from 3.1 to 3.3 (Fig. 13). This also shows the applicability of the proposed anode gas diffusion resistance to AC impedance measured at different temperatures.

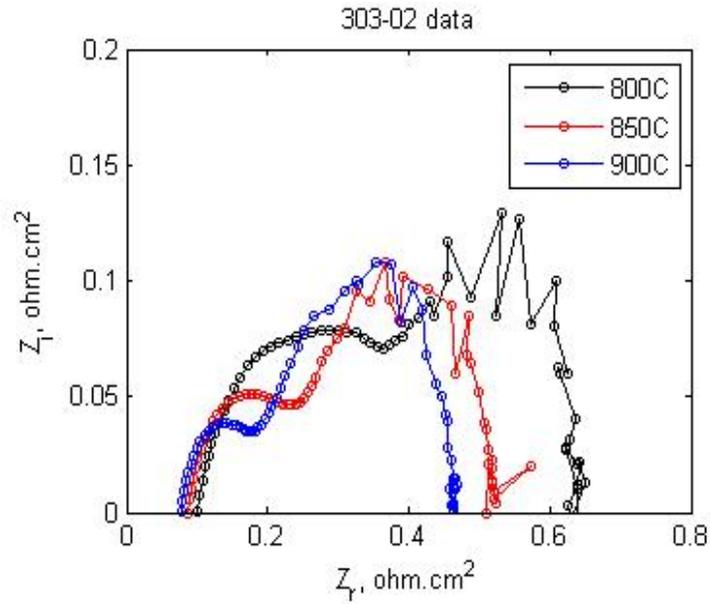

Fig. 12 OCV AC impedance data of anode supported cell collected at various temperatures. Hydrogen partial pressure is fixed at 100%, and total pressure of the anode feeding gas was fixed at 1 atm.

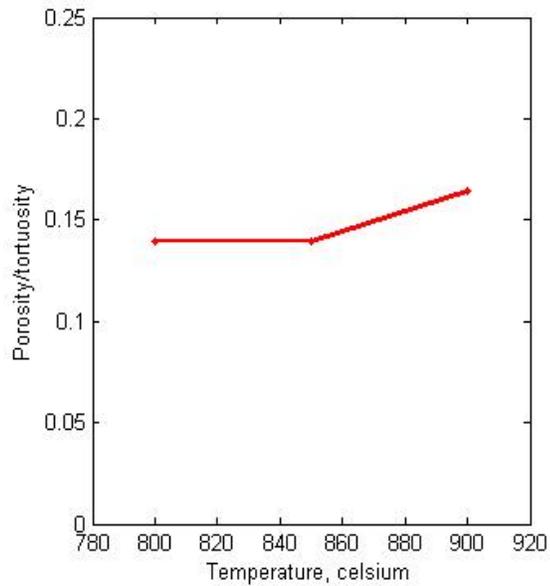

Fig. 13 Comparison of the fitted structure factor at three different temperatures. (800ºC, 850ºC and 900ºC)

V. **Discussion and Conclusion**

In this work, we investigated the multi-component gas transport in porous electrodes using anode-supported solid oxide fuel cells, and proposed a new theoretical approach to predict gas diffusion resistance ($R_b$). Explicit analytical expressions for gas diffusion resistance ($R_b$) were derived at zero current conditions, and values of gas diffusion resistance ($R_b$) were evaluated numerically at non-zero current conditions. Comparison of cathode and anode gas diffusion resistance shows that in anode supported cell, anode is the major contributor to gas diffusion resistance.

Experimental $R_b$ values were determined by fitting the low frequency arc of the anode supported cell to the finite length Warburg impedance in a Randles circuit. Then, they were compared to predictions from three analytical models, including Fick's Law, Stefan-Maxwell model, and Dusty Gas model, to determine the structure factor (porosity divided by tortuosity) or tortuosity when porosity is known. An inconsistency between isobaric assumption and the Dusty Gas model was identified, but numerical simulation confirmed that total pressure variation only has very small effects on gas composition profiles and the predicted gas diffusion resistance ($R_b$). Therefore, we can still safely use isobaric assumptions with Dusty Gas model. By incorporating interactions between different gas molecules and collisions between gas molecules and pore walls (Knudsen effect), Dusty Gas model works best and gives a more or less constant tortuosity value over a wide range of operating conditions (10% to ~100% of hydrogen partial pressure, zero and non-zero currents, and three different temperatures), and the fitted tortuosity value matches well with direct experimental measurements.

In summary, this work developed a new theoretical approach to utilize AC impedance data and various analytical models to investigate multicomponent gas diffusion in porous media. The remarkable data collapse of the measured gas diffusion resistance for a wide range of hydrogen partial pressures, currents and temperatures with single, reasonable tortuosity establishes DGM as the best model for gas diffusion in porous media (at least under these conditions). Therefore, this approach can be used to estimate tortuosity for porous media or to estimate gas diffusion resistance for further investigating other physical processes occurring inside the porous electrodes.

This study further shows that electrochemical impedance analysis is a much more reliable method to obtain gas diffusion information for porous media than other methods based on permeability or limiting-current measurements.